\documentclass[12pt,preprint]{aastex} 

\begin{document}

%%%%%%%%%%%%%% MY DEFINITIONS
\def\agile {\emph{AGILE}}
\def\xmm {\emph{XMM-Newton}}
\def\cha {\emph{Chandra}}
\def\comp {\emph{COMPTEL}}
\def\flux {\mbox{erg cm$^{-2}$ s$^{-1}$}}
\def\lum {\mbox{erg s$^{-1}$}}
\def\nh {$N_{\rm H}$}
\def\psr {PSR~B1509--58}

%%%%%%%%%%%%%%%%%%%%%%%%%%%%%%%%%%%%%%

\title{\emph{AGILE} OBSERVATIONS OF THE ``SOFT" GAMMA-RAY PULSAR PSR~B1509--58}

\shorttitle{1509}
\shortauthors{M.~Pilia et al.}

\author{M.~Pilia\altaffilmark{1,2}, A.~Pellizzoni\altaffilmark{2}, 
A.~Trois\altaffilmark{3}, F.~Verrecchia\altaffilmark{4}, 
P.~Esposito\altaffilmark{2,6},
P.~Weltevrede\altaffilmark{7,8}, 
S.~Johnston\altaffilmark{7},
M.~Burgay\altaffilmark{2},
A.~Possenti\altaffilmark{2},
E.~Del Monte\altaffilmark{3},
F.~Fuschino\altaffilmark{9},
P.~Santolamazza\altaffilmark{4}, 
A.~Chen\altaffilmark{5,10}, 
A.~Giuliani\altaffilmark{5}, 
P.~Caraveo\altaffilmark{5}, 
S.~Mereghetti\altaffilmark{5}, M.~Tavani\altaffilmark{3,11}, 
A.~Argan\altaffilmark{3}, 
E.~Costa\altaffilmark{3},
N.~D'Amico\altaffilmark{2},
A.~De Luca\altaffilmark{5,6,12}, 
Y.~Evangelista\altaffilmark{3}, M.~Feroci\altaffilmark{3}, 
F.~Longo\altaffilmark{13}, M.~Marisaldi\altaffilmark{9}, 
G.~Barbiellini\altaffilmark{13}, 
A.~Bulgarelli\altaffilmark{9}, 
P. W. Cattaneo\altaffilmark{6}, V.~Cocco\altaffilmark{3}, 
F.~D'Ammando\altaffilmark{3,11,14}, G.~De Paris\altaffilmark{3}, 
G.~Di Cocco\altaffilmark{9}, I.~Donnarumma\altaffilmark{3}, 
M.~Fiorini\altaffilmark{5}, T.~Froysland\altaffilmark{10,11}, 
M.~Galli\altaffilmark{15}, F.~Gianotti\altaffilmark{9}, 
C.~Labanti\altaffilmark{9}, I.~Lapshov\altaffilmark{3}, 
F.~Lazzarotto\altaffilmark{3}, P.~Lipari\altaffilmark{16}, 
A.~Morselli\altaffilmark{17}, 
L.~Pacciani\altaffilmark{3}, F.~Perotti\altaffilmark{5}, 
G.~Piano\altaffilmark{3,11},
P.~Picozza\altaffilmark{17}, M.~Prest\altaffilmark{1}, 
G.~Pucella\altaffilmark{18}, M.~Rapisarda\altaffilmark{18}, 
A. Rappoldi\altaffilmark{6},
S.~Sabatini\altaffilmark{3,16},
P.~Soffitta\altaffilmark{3}, 
M.~Trifoglio\altaffilmark{9}, E.~Vallazza\altaffilmark{13}, 
S.~Vercellone\altaffilmark{14}, V.~Vittorini\altaffilmark{11}, 
A.~Zambra\altaffilmark{5}, D.~Zanello\altaffilmark{16}, 
C.~Pittori\altaffilmark{4}, 
F.~Lucarelli\altaffilmark{4},
P.~Giommi\altaffilmark{4}, L.~Salotti\altaffilmark{19}, and
G.~F.~Bignami\altaffilmark{12}}

\altaffiltext{1}{Dipartimento di Fisica, Universit\`a dell'Insubria, Via Valleggio 11, I-22100 Como, Italy}
\altaffiltext{2}{INAF--Osservatorio Astronomico di Cagliari,
  localit\`a Poggio dei Pini, strada 54, I-09012 Capoterra, Italy}
\altaffiltext{3}{INAF/IASF--Roma, Via del Fosso del Cavaliere 100,
  I-00133 Roma, Italy}
\altaffiltext{4}{ASI--ASDC, Via G. Galilei, I-00044 Frascati (Roma), Italy}
\altaffiltext{5}{INAF/IASF--Milano, Via E.~Bassini 15, I-20133 Milano,
  Italy}
\altaffiltext{6}{INFN--Pavia, Via Bassi 6, I-27100 Pavia, Italy}
\altaffiltext{7}{Australia Telescope National Facility, CSIRO, P.O.~Box~76,
  Epping NSW~1710, Australia}
\altaffiltext{8}{Jodrell Bank Centre for Astrophysics, The Alan Turing Building, School of
Physics and Astronomy, The University of Manchester, Oxford Rd, Manchester,
M13 9PL, UK}
\altaffiltext{9}{INAF/IASF--Bologna, Via Gobetti 101, I-40129 Bologna,
  Italy}
\altaffiltext{10}{CIFS--Torino, Viale Settimio Severo 3, I-10133, Torino, Italy}
\altaffiltext{11}{Dipartimento di Fisica, Universit\`a ``Tor Vergata'', Via della Ricerca Scientifica 1, I-00133 Roma, Italy}
\altaffiltext{12}{Istituto Universitario di Studi Superiori, V.le Lungo Ticino Sforza 56, 27100 Pavia, Italy}
\altaffiltext{13}{Dipartimento di Fisica, Universit\`a di Trieste and
  INFN--Trieste, Via Valerio 2, I-34127 Trieste, Italy}
\altaffiltext{14}{INAF/IASF--Palermo via U. La Malfa 153, I-90146,
  Palermo, Italy}
\altaffiltext{15}{ENEA--Bologna, Via Biancafarina 2521, I-40059
  Medicina (BO), Italy}
\altaffiltext{16}{INFN--Roma ``La Sapienza'', Piazzale A. Moro 2, I-00185 Roma, Italy}
\altaffiltext{17}{INFN--Roma ``Tor Vergata'', Via della Ricerca Scientifica 1, I-00133 Roma, Italy}
\altaffiltext{18}{ENEA--Roma, Via E. Fermi 45, I-00044 Frascati (Roma), Italy}
\altaffiltext{19}{ASI, Viale Liegi 26 , I-00198 Roma, Italy}

\email{mpilia@ca.astro.it}

\begin{abstract}

We present the results of new \agile\ observations of \psr\
performed over a period of $\sim$2.5 years
following the detection obtained with a subset of the present data. 
The modulation significance of the lightcurve above 30 MeV is at a 
5$\sigma$ confidence level and the lightcurve is similar to those 
found earlier by \comp\ up to 30 MeV: a broad asymmetric first peak reaching 
its maximum $0.39 \pm 0.02$ cycles after the radio peak plus a second peak at
$0.94 \pm 0.03$.  
The gamma-ray spectral energy distribution of the pulsed flux detected by
\comp\ and \agile\ is well
described by a power-law (photon index $\alpha=1.87\pm0.09$) with a remarkable 
cutoff at $E_c=81\pm 20$~MeV, representing the softest
spectrum observed among gamma-ray pulsars so far.
The pulsar luminosity at $E > 1$~MeV is $L_{\gamma}=4.2^{+0.5}_{-0.2} \times
10^{35}$~erg/s, assuming a distance of 5.2~kpc,
which implies a spin-down conversion efficiency to gamma-rays of $\sim 0.03$.
The unusual soft break in the spectrum of PSR B1509-58 has been interpreted 
in the framework of polar cap models as a signature of the exotic photon
splitting process 
in the strong magnetic field of this pulsar. In this interpretation our
spectrum  
constrains the magnetic altitude of the emission point(s) at 3~km
above the neutron 
star surface, implying that the attenuation may not be as strong as formerly
suggested because pair production can substitute photon splitting in regions of
the magnetosphere where the magnetic field becomes too low to sustain photon
splitting.  
In the case of an outer-gap scenario, or the two pole caustic model, better
constraints on the geometry of the emission would be needed from the radio
band in 
order to establish whether the conditions required by the models to reproduce
\agile\ lightcurves and spectra match the polarization measurements.

\end{abstract}


\section{Introduction}

\psr\  (J1513-5908) was discovered as an X-ray pulsar with the {\it Einstein} satellite during an observation 
of the supernova remnant (SNR) MSH~15-52 \citep{seward82}. The  source  was  soon also detected at radio frequencies by
\cite{manchester82},  with  a  derived distance
supporting the  association with  the SNR ($d\sim
5.2$~kpc, as calculated from HI measurements from \citealt{gaensler99} and in
agreement with the most recent distance derived using the dispersion measure \footnote{see the ATNF Pulsar
Catalogue (http://www.atnf.csiro.au/research/pulsar/psrcat/) for the updated
distance measurement derived from the dispersion measure.}).  
With a period $P\simeq  150$~ms and a period derivative
$\dot P  \simeq 1.53 \times 10^{-12}$~s~s$^{-1}$,  assuming the standard dipole
vacuum model, the estimated spin-down 
age for  this pulsar is 1570  years (among the shortest for radio pulsars)
and  its inferred  surface
magnetic field is one of the  highest observed for an ordinary radio pulsar: $B=
3.1\times 10^{13}$~G, as calculated at the pole
\footnote{The magnetic
field strength at the pole is twice the value quoted in the ATNF Pulsar
Catalogue.}.
Its rotational energy  loss rate is $\dot E =
1.8 \times  10^{37}$~erg/s. 

\psr\  and its  nebula have been extensively  observed  in the
X-ray energies since the Eighties with
the  {\it Einstein} and {\it EXOSAT} satellites.
The detection  of pulsed emission in  the hard X-rays dates back to the
early Nineties  \citep{kawai91} with {\it Ginga}
in the 2--60~keV energy range.
During a 20~yrs-long radio monitoring 
\citep{livingstone05}, \psr\ has
not  shown any  glitch activity, at variance with the general behavior 
of young  radio pulsars, which usually show some glitch activity.
The analysis of \cite{livingstone05}, using radio and X-rays (collected with
{\it RossiXTE}), yielded a very accurate measurement of the braking  index  
of  $n=2.839\pm 0.003$, close  to the
canonical  value $n=3$  for braking  by magnetic  dipole  radiation in
vacuum alone.
Observations with the {\it ROSAT} \citep{trussoni96}, {\it ASCA}
\citep{tamura96}  and {\it BeppoSAX} 
\citep{mineo01} satellites 
were performed in the
Nineties, characterizing the spectrum of the pulsed emission and the
morphology of the remnant as possibly due 
to  the  presence  of  several components, interacting  via
collimated outflows  from the pulsar.
The nebula has been extensively observed with the {\it Chandra} satellite
\citep{gaensler02} 
and its emission has been found up to the TeV energies, with
{\it CANGAROO} first \citep{sako00} and more recently by {\it H.E.S.S.}
\citep{aharonian05msh}. 

The young age  and the  high rotational  energy loss rate made this
pulsar a promising target for the first generation of gamma-ray
satellites. In fact, the instruments on the
{\it Compton Gamma-Ray Observatory} ({\it CGRO}) observed its
pulsation at low gamma-ray energies: up to $E\sim700$~keV with {\it BATSE}
\citep{wilson93b} 
and  {\it OSSE} (\citealt{ulmer93},  \citealt{matz94}), and  in the
0.75~-~30~MeV band with \comp\ \citep{kuiper99}, but it was not detected with
high 
significance by the {\it Energetic Gamma-Ray Experiment Telescope} ({\it
  EGRET}), 
the instrument  operating at  the energies from 30~MeV to 30~GeV.
This was remarkable, since all other known gamma-ray pulsars show spectral
turnovers well above 100~MeV \citep{thompson04}.  
\cite{harding97} suggested that the break in the spectrum  
could be interpreted as due to inhibition of the pair-production
caused by the photon-splitting phenomenon \citep{adler70}. The photon
splitting appears, in the frame of the polar cap models, in
relation with a very high magnetic field. 
An alternative explanation is proposed by \cite{zhangcheng00} using a
three dimensional outer gap model. They propose that the gamma-ray emission is
produced by synchrotron-self Compton radiation above the outer gap.

Ten  years after  the {\it CGRO},  the observation in the gamma-ray  band are
  possible again with  the  advent   of  two  high-energy  missions: the {\it
  Astro-rivelatore  
Gamma a Immagini LEggero} (\agile) and {\it Fermi} satellites.
With its large field of view 
($\sim 2.5$~sr) and its $\lesssim 1\mu$s time tagging capability
\citep{tavani09} 
(with $200\mu$s absolute timing accuracy, see \citealt{pellizzoni09a}), \agile\
is very 
well suited for the
observation of pulsars between 30~MeV and a few GeV with its {\it Gamma-Ray
  Imaging Detector} ({\it GRID}).
In particular, despite its lower sensitivity in the GeV band, the 
{\it GRID} on board \agile\ has 
an effective area below 100 MeV ($\sim$200 cm$^2$ at 50 MeV) comparable 
with that of {\it Fermi}.
\agile\ obtained the first detection of \psr\ in the {\it EGRET} band
\citep{pellizzoni09b}  confirming the 
occurrence of a spectral break, 
although no precise flux measurements were possible due to low counts
statistics. 
Recently {\it Fermi} also reported its detection of \psr\ \citep{abdo10_1509}.
In this paper we present the results of a $\sim 2.5$~yr monitoring campaign
of \psr\ with \agile, that improved counts statistics, and therefore
lightcurve characterization, with respect to earlier \agile\ observations. 
With these observations the spectral energy distribution (SED) at energies
$E<300$~MeV, where the remarkable spectral turnover is
observed, can be assessed.

\section{\emph{AGILE} Observations, Data Analysis and Results}

\psr\ is within the same region of the sky as the Vela
pulsar, an area to which 
\agile\ devoted a large amount of observing time 
(for details on \agile\ observing strategy, timing calibration and gamma-ray
pulsars analysis see \citealt{pellizzoni09a,pellizzoni09b}).
Gamma-ray photons for this pulsar were collected and analyzed starting in
July 2007, up to late October 2009 when \agile\
started observing in spinning mode due to reaction wheel 
failure
\footnote{This failure is not affecting AGILE/GRID sensitivity and pulsar
observations although the new spinning mode required calibration
revisions that are still ongoing.}. 
The large \agile\ effective area and long observing time ($\sim 260$ days on
target) provide a total exposure of 
$3.8\times 10^{9}$~cm$^2$~s ($E > 100$ MeV) during this $2.5$~yr period 
which gives our observations a good
photon harvest from this pulsar.

\subsection{Timing Analysis}

Simultaneous radio observations of \psr\ with the 64~m
Parkes radiotelescope in Australia are ongoing since the epoch of
\agile's launch (MJD 54220; 2007 April 30), as part of a timing project for
the gamma-ray satellites \citep{weltevrede10timfermi}, and cover all of \agile's
observations. 
 A total of 47 pulsar  time of arrivals (ToAs) were collected between April 2007 (MJD 54220) and February 2010 (MJD 55233), leading to a
 r.m.s. of the residuals of $900$~$\mu$s, 
showing the goodness of the timing model that
allowed accurate pulse phase tagging of the gamma-ray photons.
No glitch was detected in the radio analysis. Strong timing noise was
present, as expected from a young pulsar, and it was accounted for using the
$fitwaves$ technique developed in the framework of the TEMPO2 radio pulsar
timing software \citep{hobbs04, hobbs06}.   
Using the radio ephemeris provided by the Parkes telescope, 
we performed the folding of the gamma-ray lightcurve including the wave terms
(see \citealt{pellizzoni09a}). 
An optimized analysis followed, 
aimed at cross-checking and maximization 
of the significance of the detection, including an energy-dependent event 
extraction angle around the source position based on the instrument
point-spread-function (PSF). 
Only high confidence 
gamma-ray photons (G) were used for the timing analysis of
this pulsar. 
The chi-squared ($\chi^2$)-test applied to the 10 
bin lightcurve at $E>30$ MeV  gave a detection significance of $\sigma = 4.8$.
The unbinned $Z_n^2$-test applied to the photons' arrival times
gave a significance of $\sigma = 5.0$ with $n=2$ harmonics. 
The difference between the radio and gamma-ray ephemerides was 
$\Delta P _{radio,\gamma}\lesssim 10^{-9}$~s, well within the error
in the parameters, showing perfect agreement
among radio and gamma-ray ephemerides as
expected, further supporting our detection and \agile\ timing calibration.

We observed \psr\ in three energy bands. 
We obtained 
$1210 \pm 400$ pulsed counts ($\sim 5$\% of the total
source, diffuse gamma-ray emission and residual particle background
counts) at energies $30<E<100$~MeV, $820 \pm 360$ pulsed counts ($\sim 7$\% of the
total counts)
at energies $100<E<500$~MeV.
The pulsed flux was computed considering all the counts above the minimum of
the lightcurve (see \citealt{pellizzoni09a})
We did not detect pulsed emission at a significance $\sigma \geq 2$ for $E > 500$~MeV
 and thus we can only give an upper limit at 1$\sigma$ of $< 270$ pulsed counts.
This is consistent with the fact that \cite{abdo10_1509} report only a
$1.4\sigma$ detection at $0.3<E<1$~GeV with the {\it Fermi} data. 

\begin{figure}
%\resizebox{\hsize}{!}{\includegraphics[angle=0]{figure1.ps}}
\centering
\includegraphics[width=0.8\textwidth]{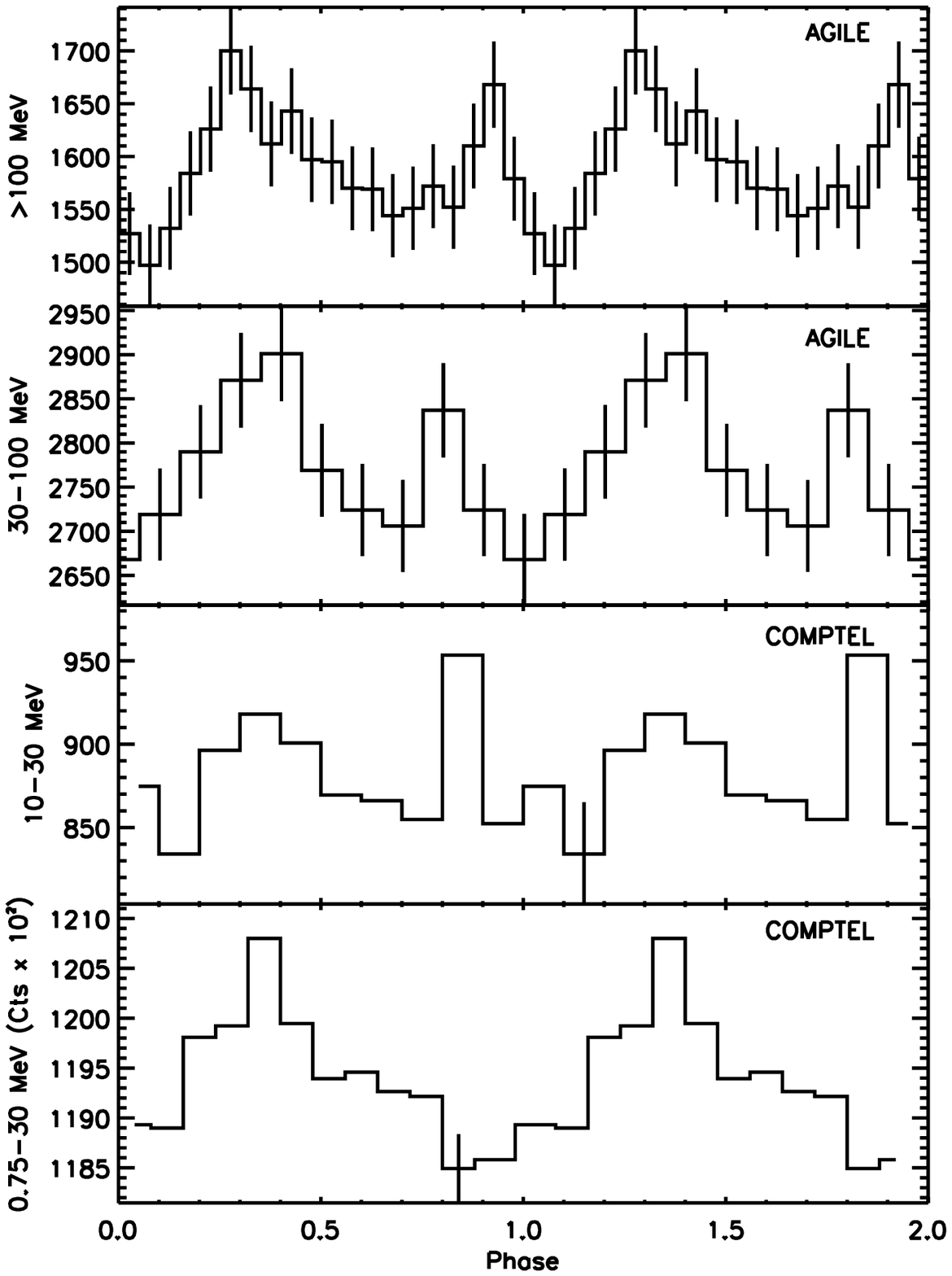}
\caption{\label{fig:lc_tot}
Phase-aligned gamma-ray light-curves of \psr. Radio main peak is at
phase 0. The start of the y-axis coincides with the minimum of the pulsed
fraction and, consequently, with the
background level.
From top to bottom: \agile\ high energy band
  ($>100$~MeV), 20 bins, 7.5
ms resolution; \agile\ ``soft'' energy band ($<100$~MeV), 10 bins, 15 ms
resolution; 
\comp\ high energy band (10--30~MeV) and
\comp\ whole bandwith (0.75--30~MeV) (from \citealt{kuiper99}).}
\end{figure}

The gamma-ray lightcurves of \psr\ for different energy bands
are shown in Fig. \ref{fig:lc_tot}. 
The \agile\ lightcurve above 30~MeV shows two peaks at phases
$\phi_1  = 0.39 \pm 0.02$ and $\phi_2  = 0.94 \pm 0.03 $ with respect to the single radio peak, 
here put at phase 0, as obtained from the Parkes ephemeris. 
The peak positions and widths in term of phase are calculated using a Gaussian fit, yielding
a FWHM of $0.29(6)$ for the first peak and of $0.13(7)$ for the second peak,
where we quote in parentheses (here and throughout the paper)
the 1$\sigma$ error on the last digit. The errors considered are statistical,
as the systematic errors do not affect the measurements of the pulsed counts.
The first peak is coincident in phase
with the X-ray single broad peak and with \comp\ peak (see \citealt{kuiper99}
and references therein). In its 
highest energy band (10--30~MeV) \comp\ showed the hint of a second peak 
(even though the modulation had low significance, $2.1 \sigma$), which
is also visible in the lightcurve derived from {\it EGRET}
observations (30--100~MeV), 
despite the fact that {\it EGRET} did not have a significant
detection of the pulsar above 100~MeV \citep{fierro95}.
This second
peak is coincident in phase with \agile's second peak
(Fig. \ref{fig:lc_tot}) in its lower energy band while it appears
slightly shifted at energies above 100 MeV.
A possible explanation for this shift is
discussed in section $\S$ 3.
\agile\ thus confirms the previously marginal detection of a second
peak, statistically significant at $5 \sigma$, calculated using
a chi squared statistics test.

\subsection {Spectral Analysis, Spectral Energy Distribution}

Based on our exposure, calculated by the {\it GRID} scientific analysis task
AG\_ExmapGen, we derived the gamma-ray flux from the number of
pulsed counts. This method, though typically giving higher statistical errors than the likelihood
analysis, is more accurate and sensitive to evaluate the flux of this pulsar, 
given its soft spectrum (and the correspondingly large PSF) and the 
contribution from other nearby and brighter sources and possibly from the
pulsar wind nebula (PWN),
that all affect the spatial analysis.
Using this method, the problem of modeling the background is dealt with by
discarding the counts below the pulsed threshold, so that the observed pulsed
counts belong to the pulsar. All the other sources of systematical errors, such as the effective area and  \agile's PSF, as well as the diffuse emission, contribute to $<10$\%, so that they are much lower than the statistical errors and they are not quoted in the flux measurements.  
We divided \agile\ bandwidth into three energy intervals: 30--100~MeV,
100--500~MeV and above 500~MeV.
The pulsed fluxes thus obtained were $F_{\gamma}= 10(3)\times 10^{-7}
$~ph~cm$^{-2}$~s$^{-1}$ in the 30--100~MeV band, 
$F_{\gamma}= 1.8(8)\times 10^{-7} $~ph~cm$^{-2}$~s$^{-1}$ in the 100--500~MeV
band  
and a $1 \sigma$ upper limit $F_{\gamma}< 8\times 10^{-8}
$~ph~cm$^{-2}$~s$^{-1}$ for $E>500$~MeV.
Finally, from the total number of pulsed counts we obtained a pulsed flux at
$E>30$~MeV 
$F_{\gamma}= 12(2)\times 10^{-7} $~ph~cm$^{-2}$~s$^{-1}$ for $E>30$~MeV.

\begin{figure}
\resizebox{\hsize}{!}{\includegraphics[angle=0]{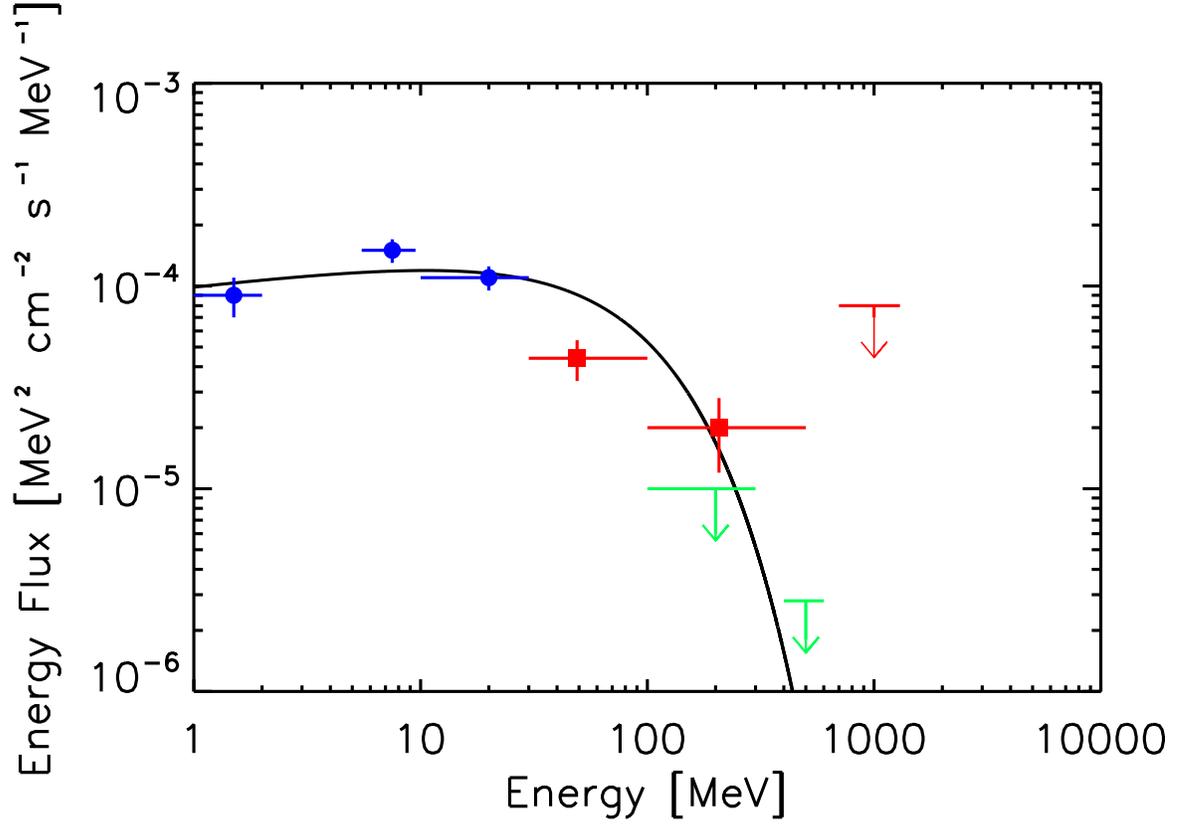}}
\caption{\label{spec} Spectral energy distribution of \psr\ (solid
  line) obtained from a fit of pulsed fluxes from soft to hard
  gamma rays. The three round
   points represent \comp\ observations \citep{kuiper99}. The two square points
  represent 
  \agile\ pulsed flux in two bands ($30<E<100$~MeV and $100<E<500$~MeV). 
The red horizontal bar and arrow emerging from it represent
  \agile\ upper limit above 500~MeV.  The two green arrows represent {\it
  Fermi} upper limits \citep{abdo10_1509}
} 
\end{figure}

Fig. 2 shows the SED of \psr\ based on
\agile's and \comp's observed fluxes. {\it Fermi} upper limits are also shown,
which are consistent with our measurements at a $2\sigma$ confidence level.
\comp\ observed this pulsar in three energy bands: 0.75--3~MeV,
3--10~MeV, 10--30~MeV, suggesting a 
spectral break between 10 and 30 MeV. 
\agile\ pulsed flux 
confirms the presence of a soft spectral break.
As shown in Fig. \ref{spec}, we modeled
the observed \comp\ and \agile\ fluxes with a power-law plus cutoff fit 
using the Minuit minimization package \citep{james75}: $F(E)=k \times
E^{-\alpha}\exp[-(E/E_{c})^{\beta}]$, 
with three free parameters: the normalization $k$, the spectral index
$\alpha$, the cutoff energy $E_c$ and allowing $\beta$ to assume values of 1
and 2 (indicating either an exponential or a superexponential cutoff). 
No acceptable $\chi^2$ values were obtained for a superexponential cutoff, the
presence of which can be excluded at a
$3.5\sigma$ confidence level,
while for an exponential cutoff we found $\chi^2_{\nu}=3.2$ for $\nu = 2$
degrees of freedom, corresponding to a
null hypothesis probability of 0.05. 
The best values thus obtained for the parameters of the fit were:
$k=1.0(2)\times 10^{-4}$~s$^{-1}$~cm$^{-2}$, $\alpha=1.87(9)$, $E_{c}=81(20)$~MeV.

We performed an analysis of the ratio between the two peak heights. 
The second peak appears in the \comp\ 
band 10--30 MeV and is observed with \agile\ up to $E\lesssim 500$~MeV: it is harder than the first
peak in the \comp\ energy band, 
and it is present at all energies in the \agile\ energy band,
so that it might possibly be harder even at \agile's energies
but the low statistics at high energies do not allow us to discriminate.

As a consistency check for the pulsed fluxes reported above, 
a maximum likelihood analysis in a region of 10 degrees around the source
position was performed to assess possible 
unpulsed contribution from the PWN
\footnote{See e.g. \cite{pellizzoni10} as an example of study of gamma-ray PWN
with \agile.},
although no detection was reported in the First Catalog of High-Confidence 
Gamma-ray Sources detected by the \agile\ satellite \citep{pittori09}. 
The likelihood analysis (see \citealt{mattox96} and for Agile in particular, details will be
provided in Chen et al. in preparation) took into account the numerous sources
present in this 
crowded region (including the extremely bright nearby gamma-ray pulsar
J1509-5850, \citealt{weltevrede10fermi}).
The upper limit found in the \agile\ 
energy range by likelihood analysis ($F_{\gamma}<40\times 10^{-8}$~ph~cm$^{-2}$~s$^{-1}$
above 100~MeV) is above 
the corresponding pulsed flux above 100~MeV 
($F_{\gamma}=21(6)\times 10^{-8}$~ph~cm$^{-2}$~s$^{-1}$). 
This is compatible with the  
fact that the timing analisys is expected to have for this target a 
better sensitivity
(with respect to the likelihood analysis).
It is worth noting that \psr\ is also not detected by likelihood analysis 
by {\it Fermi} \citep{abdo10_1509} apart from the $>1$~GeV energy band 
where the emission could be related to the PWN seen by {\it H.E.S.S.} 
(\citealt{aharonian05msh}).

\section{Discussion}

Pulsar magnetosphere models are usually divided into two categories, depending
on the sites for the high-energy emission. In {\it polar cap} models
\citep{dauhar96} the
emission comes from the regions near the neutron star surface, while {\it
  outer gap} models \citep{chr86,romani96} predict that the emission be
originated in the regions 
close to the light cylinder. Alternative models predict an emission zone
encompassing the whole magnetosphere, which departs from the external rim of
the polar cap region: these are the {\it slot gap} models \citep{mushar03};
others predict emission from alternative regions in the outer magnetosphere,
the {\it annular gap}
(\citealt{du10} and references therein).
Different models predict different spectral and geometrical properties.
The bulk of the spin-powered pulsar flux is usually emitted in the MeV-GeV
energy band with  
spectral breaks at $\lesssim 10$~GeV (see \citealt{abdo10psrcat} or
e.g. \citealt{aliu08crab}).
\psr\ has the softest spectrum observed among gamma-ray
pulsars, with a sub-GeV cutoff at $E = 0.08(2) $~GeV. 
The second softest spectrum
and lowest energy cutoff (0.7(5)~GeV) is that of PSR~B0656+14, recently
observed 
by {\it Fermi} \citep{weltevrede10fermi}.
The observed lightcurve of \psr\ shows two peaks lagging the radio peak by,
respectively, $\phi_1  = 0.39 \pm 0.02$ and $\phi_2  = 0.94 \pm 0.03 $. 
In the following we discuss how the new \agile\ observations can constrain the
models for emission from the pulsar magnetosphere.

When \psr\ was detected in soft gamma-rays but not significantly at $E>30$~MeV,
it was proposed that the mechanism 
responsible for this low-energy spectral break might be photon splitting
\citep{harding97}.
The photon splitting \citep{adler70} is an exotic third-order quantum
electro-dynamics (QED) process expected when the 
magnetic field approaches or exceeds the $critical$ value defined as
$B_{cr}=m^2_e c^3/(e\hbar)=4.413\times 10^{13}$~G, above which quantum effects become relevant. 
Most current theories for
the generation of coherent radio 
emission in pulsar magnetospheres require formation of an
electron-positron pair plasma developing via electromagnetic cascades. In very
high magnetic fields the formation of pair cascades can be altered
by the process of photon splitting: $\gamma \rightarrow \gamma\gamma$, 
which will operate as an 
attenuation mechanism in the high-field regions near pulsar polar caps. 
Since it has no energy threshold, photon splitting can attenuate photons below
the threshold for pair production, 
thus determining a spectral cutoff at lower energies.
This process cannot operate in the low fields of outer gap models
because it only has appreciable reaction rates when the magnetic field is at
least a significant fraction of the quantum critical field $B_{cr}$
(the attenuation coefficient $T_{sp}$ scaling as 
$T_{sp}\propto (B/B_{cr})^{6}=B^{\prime 6}$), and magnetic
fields strong enough are not present in the outer magnetosphere as $B \sim
r^{-3}$.  

In the case of \psr\ a polar cap model with photon splitting would be
able to explain the soft gamma-ray emission and the low energy
spectral cutoff, now quantified by \agile\ observations.
Since the mechanism of photon splitting is, as stated, strongly dependent
on the magnetic field strength,
if the field strength at the emitting region is
$B^{\prime} \gtrsim 0.3$
(i.e. at heights below 1.3 neutron star radii, $R_{NS}$), the photon splitting
is the dominant means 
of attenuation
that inhibits efficient pair cascade production \citep{harding97} and then
gamma-ray emission.
Based on the observed cutoffs, which are related to the photons' saturation
escape energy, 
we can derive constraints on the magnetic field strength at emission,
in the framework of photon splitting:

\begin{equation}
\epsilon_{esc}^{sat} \simeq 0.077(B^{\prime}  \sin \theta_{kB,0})^{-6/5} 
\label{eq:emax}
\end{equation}

where $\epsilon_{esc}^{sat}$ is the photon saturation escape energy and
$\theta_{kB,0} $ is the angle between the 
photon momentum and the magnetic field vectors at the surface and is here
assumed to be very small: 
 $\theta_{kB,0} \lesssim 0.57 ^{\circ} $
(see \citealt{harding97}). 
Using the observed energy cutoff ($\epsilon_{esc}^{sat} \simeq E= 80$~MeV) we
find that $B^{\prime} \gtrsim 0.3$, which 
implies an emission altitude $\lesssim 1.3 R_{NS}$, which is the height where
possibly also pair production could ensue. 
This altitude of emission agrees with the polar cap models
(see e.g. \citealt{dauhar96}). A smaller energy cutoff, as in
\cite{harding97}, would have implied even lower emission altitude and a
sharper break, possibly caused by the total absence of pair production. 
It is apparent that small differences in the emission position will cause
strong differences in  spectral shape. This is possibly the reason for 
the different emission properties of the two peaks as observed in the total
(\agile\ plus \comp)  
gamma-ray energy band. Also, a trend can be observed, from lower to higher
energies (see the X-ray lightcurve for the trend in the first peak, as in
Fig. 3 of \citealt{kuiper99}), of the peaks slightly drifting away from the
radio peak. This we assume 
to be another signature of the fact that small variations in emission height
can be responsible for sensible changes in the lightcurves
in such a high magnetic field.
The scenario proposed by \cite{harding97} 
is strengthened by its prediction that PSR~B0656+14
should have a cutoff with an intermediate value between \psr\ and the
other gamma-ray pulsars. 
The main reason for the parallel between the two pulsars was at the time the
fact that they had, respectively, the highest and second highest inferred
magnetic fields. At present, however, there are a handful of gamma-ray pulsars
with higher magnetic field than PSR~B0656+14
in the {\it Fermi} First Year Pulsar
Catalog \citep{abdo10psrcat} 
which do not show a low energy cutoff. 
Nonetheless, \psr\ (see \citealt{kuiper99,crawford01}) and PSR~B0656+14 
\citep{deluca05,weltevrede10fermi} both show
evidence of being aligned rotators, which could imply polar cap emission, as
is also hinted by \cite{b&s10}.  

A soft cutoff (below 1~GeV) is in principle possible for polar cap 
scenarios even without invoking photon splitting attenuation. In
polar cap models the strong magnetic field permits one-photon pair creation
that 
attenuates super-GeV photons in Crab-like (e.g. \psr, based on the
parameter $B/P^2$ ) and Vela-like pulsars (e.g. PSR~B0656+14), whereas pair
creation in outer gap models is mediated through the two-photon process
involving surface thermal X-rays as targets. According to the calculations of
\cite{dauhar96}, emission from the regions close to the polar caps
is possible when $\alpha \sim \theta_{b}$, where $\alpha$ is the angle between
  the rotation and the magnetic axis and $\theta_{b}$ is the half-angle of the
  gamma-beam emerging from the polar cap. Furthermore, with emission from
  the polar caps, 
  or some ($\geq 2$) polar cap radii, the pulse profile at high energies can
  have either one (as
  in the case of  PSR~B0656+14) or two peaks, with a peak-to-peak phase
  separation as large as 0.4--0.5 (albeit slightly smaller than what is observed for \psr\ at the highest energies).

The polar cap model 
as an emission mechanism is nowadays debated. On one hand theoretical
objections 
arise from the fact that the angular momentum is
not conserved in polar cap emission (see \citealt{cohentreves72},
\citealt{holloway77},  
Treves et al. in preparation).
At the same time, mounting evidence of a preferential explanation of the
observed gamma-ray 
lightcurves with high altitude cascades is also coming from the recent
results by the {\it Fermi} satellite (see e.g. \citealt{abdo10psrcat}).
In the case of \psr, 
the derived gamma-ray luminosity from the flux at 
$E>1$~MeV, considering a 1~sr beam sweep is $L_{\gamma}=4.2^{+0.5}_{-0.2}
d^2_{5.2} \times 
10^{35}$~erg/s, where $d_{5.2}$ indicates the distance in units of 5.2 
kpc.
While traditionally the beaming fraction ($f_{\Omega}$) was considered to be
the equivalent of a 1~sr sweep, nowadays (see e.g. \citealt{watters09})
the tendency is to consider a larger beaming
fraction ($f_{\Omega} \approx 1$), close to a $4\pi$~sr beam. 
Using $f_{\Omega}=1$ in our calculations,
we would have obtained $L_{\gamma}=5.8^{+0.1}_{-0.8}d^2_{5.2}\times
10^{36}$~\lum. 
Thus the maximum conversion efficiency of the rotational
energy loss ($\dot E \approx 1.8 \times  10^{37}$~\lum, see $\S$~1) 
into gamma-ray luminosity is 0.3.
Our result is not easily comparable with the typical gamma-ray luminosities
above 100~MeV, 
because for \psr\ this energy band is beyond the spectral break.
Using \agile\ data alone we obtained a luminosity above 30~MeV
$L_{\gamma}=5.2(6) d^2_{5.2}\times  10^{35}$~erg/s, again for a 1~sr beam.
If the gamma-ray 
luminosity cannot account for a large fraction of the rotational energy loss,
then the 
angular momentum conservation objection from \cite{cohentreves72} becomes less
cogent for this pulsar, 
exactly as it happens for the radio emission. 
For PSR~B0656+14 no
outer magnetosphere emission model seemed to satisfy the observed features and
a lower magnetosphere model, with an aligned geometry between the rotational
and magnetic 
axes, has been proposed and 
seems plausible from polarization studies. Its efficiency in the conversion of
the rotational energy loss into gamma-ray luminosity is one of
the lowest observed for the gamma-ray pulsars (see \citealt{pellizzoni09b},
\citealt{abdo10psrcat}): $\eta=0.01$, 
not violating the constraints imposed by the conservation of angular momentum.
 
Alternatively, if such an efficiency as that of \psr\ were incompatible with
this conservation law, an interpretation of \psr\ emission should be sought
 in the frame of the three dimensional outer magnetosphere gap model, as was done
by \cite{zhangcheng00}.
According to their model, hard X-rays and low energy gamma-rays have the same
origin: they are produced by synchrotron self-Compton radiation of secondary
electron-positron pairs of the outer gap. Therefore, as observed, the phase
offset of hard X-rays and low energy gamma-rays with respect to the radio
pulse is the same, with the possibility of a small lag due to the thickness of
the emission region.
According to \cite{zhangcheng00} estimates, 
a magnetic inclination angle $\alpha\approx 60 ^o$ and a viewing
angle $\zeta \approx 75 ^o$ are
required to reproduce the observed lightcurve. Similarly, for PSR~B0656+14,
\cite{weltevrede10fermi} argue that 
large $\alpha$ and $\zeta$ angles
are required to reproduce the
observed lightcurve in the framework of outer gap models. 
Finally, using the simulations of \cite{watters09}, 
who produced a map of pulse profiles for different combinations of
angles $\alpha$ and $\zeta$ in the different emission models,
the observed lightcurve from \agile\ is best reproduced
if $\alpha\approx 35 ^{\circ}$  and   $\zeta \approx 90
^{\circ}$, in the framework of the two pole caustic model
\citep{dyksrudak03}.

Since the parameters used for the application of the outer gap model to \psr\
were based on  
its former observations by \comp, \agile\ spectrum does not precisely fit the
spectrum 
predicted by the model of \cite{zhangcheng00}. 
Furthermore, the values of $\alpha$ and $\zeta$ required by this model are not
in good 
agreement with the corresponding values obtained with radio measurements.
In fact, \cite{crawford01} observe that $\alpha$ must be $ < 60 ^{\circ}$ 
at the $3 \sigma$ confidence level.
The prediction obtained by the simulations of \cite{watters09} for slot gap
emission 
is in better agreement with the radio polarization 
observations than what predicted in the outer gap framework.
In fact, in the framework of the rotating vector model (RVM, see
e.g. \citealt{lorimer04} and references therein),
\cite{crawford01} also propose that, 
if the restriction is imposed that $\zeta > 70 ^{\circ}$ \citep{melatos97} 
then $\alpha > 30 ^{\circ}$ at the $3 \sigma$ level.
For these values, however, the Melatos model for the spin down of an oblique
rotator 
predicts a braking index $n>2.86$, slightly inconsistent with the observed 
value ($n=2.839(3)$, see \S 1).
Also in the case of
PSR~B0656+14, \cite{weltevrede10fermi} conclude that 
the large values of
$\alpha$ and $\zeta$ are somewhat at odds with the constraints from the
modeling of the radio data and the thermal X-rays which seem to imply a more
aligned geometry. 
Improved radio polarization measurements would help placing better constraints
on the pulsar geometry and therefore on the possibility of a gap in the
extended or outer magnetosphere, but the quality of the polarization
measurements from \cite{crawford01} is already excellent, 
the problem being that \psr, like most pulsars, only shows
emission over a limited pule phase range and therefore the RVM models
are highly degenerate.
At present the geometry privileged by the
state of the art measurements is best compatible with polar cap models. 
Higher statistics in the number of observed gamma-ray pulsars  could help
characterize a class of ``outliers'' having gamma-ray emission from the
polar caps, which potentially constitute a privileged target for \agile.

\section{Conclusions}

In this paper we present the result of a 2.5~yr long observation
campaign of \psr\ with \agile. 
With respect to our previous work \citep{pellizzoni09b} the increased statistics 
allowed us to perform an improved lightcurve analysis and to better constrain
the 
soft spectral cutoff observed for this pulsar.

\begin{itemize}

\item[i)]{Using the Parkes radio ephemeris, \agile\ firmly confirmed the
  detection of gamma-ray pulsation with 
good significance ($\sim 5\sigma $) from \psr.}  
\item[ii)]{The observed lightcurve shows two peaks which lag the radio peak
  of, respectively,  
$0.39\pm 0.02$ and $0.94\pm 0.03$ cycles, as obtained from a Gaussian fit of
  the peaks. 
\psr\ presents a single peaked profile up to energies $E>10$~MeV 
where \comp\ detected an additional peak with lower significance.
\agile\ confirmed the existence of this second harder peak in the 30--500 MeV
energy band.}
\item[iii)]{The detection of pulsed emission by \agile\ at $E > 30$~MeV,
  confirming  
the presence of a soft spectral break, moves the cutoff slightly up, to $E
  \approx 80$~MeV, 
in agreement with the previous estimates of a cutoff at energies below 100~MeV.
} 

\end{itemize}
Our observations are compatible with emission from the polar cap regions
powered by photon splitting cascades. 
This interpretation 
could represent a physical measurement 
related to the QED photon splitting process.
The fact that polar cap emission at high energies
appears rare might be explained by the requirement that a number of conditions
concur to have low magnetosphere emission: an aligned geometry and a high
magnetic field, without conflicting with the conservation laws.
With the \agile\ capability of observing with good sensitivity at $E>30$~MeV,
it will be 
interesting to investigate the 
highly magnetized pulsars population as a possible contributor to a new
class of ``soft'' gamma-ray pulsars. 
Alternative emission models rely on better knowledge of the geometry of \psr.

\acknowledgments
The authors thank the anonymous referee for the constructive comments.
M.P. wishes to thank Aldo Treves for useful discussion and comments
and she acknowledges the University of Insubria-Como for financial
support. 
The \agile\ Mission is funded by the Italian Space Agency (ASI) and
programmatic participation by the Italian Institute of Astrophysics (INAF)
and  the Italian Institute of Nuclear Physics (INFN).
The Parkes radiotelescope is part of the Australia Telescope, funded by
the Commonwealth Government for operation as a National Facility managed by
CSIRO.

\end{document}